\documentstyle[12pt]{article}

\newcommand{\eq}[1]{(\ref{#1})}
\newcommand{\eqs}[2]{(\ref{#1},~\ref{#2})}

\newcommand{\Eq}{Eq.~\eq}
\newcommand{\Eqs}{Eqs.~\eqs}

\newtheorem{remark}{Remark}

\newcommand{\be}{\begin{equation}}
\newcommand{\ee}{\end{equation}}
\newcommand{\ba}{\begin{eqnarray}}
\newcommand{\ea}{\end{eqnarray}}
\newcommand{\ban}{\begin{eqnarray*}}
\newcommand{\ean}{\end{eqnarray*}}
\newcommand{\ra}{\rangle}
\newcommand{\la}{\langle}

\newcommand{\k}{\kappa}

\newcommand{\A}{{\overline{A}}}
\newcommand{\w}{{\tilde{w}}}
\newcommand{\W}{{\tilde{W}}}
\newcommand{\r}{{\bf r}}
\newcommand{\cc}{{\bf c}}

\newcommand{\dd}{{\rm d}}
\newcommand{\om}{\omega}

\begin{document}

\title{Finite resolution of time \\in continuous measurements:\\
phenomenology and the model\thanks{Published in 
Physics Letters A231,  1-8 (1997)}}
\author{Michael B. Mensky\\
P.N.Lebedev Physical Institute, 117924 Moscow, Russia}
\date{}

\maketitle

\vspace{3mm}
\centerline{\large Abstract}

\begin{quote}
Definition of a quantum corridor describing monitoring a quantum
observable in the framework of the phenomenological
restricted-path-integral (RPI) approach is generalized for the case of
a finite resolution of time. The resulting evolution of the
continuously measured system cannot be presented by a differential
equation. Monitoring the position of a quantum particle is also
considered with the help of a model which takes into account a finite
resolution of time. The results based on the model are shown to
coincide with those of the phenomenological approach.
\end{quote}


\section{Introduction}

During last decades the theory of continuous quantum
measurements has been under thorough investigation both with the help
of models
\cite{cont-meas-model,CaldeiraLegg83,Milburn88,KMNamiot}
and in the framework of different phenomenological approaches
\cite{cont-meas-phenomen,M79,Mbk2,book93}
The interest to this field significantly increased in connection
with the quantum Zeno effect predicted in  \cite{Zeno} and
experimentally verified in \cite{Itano}.
Phenomenological approaches have an advantage of being
model-independent.  One of the first approaches of this type applied
to continuous quantum measurements was one based on restricted path
integrals (RPI). It was proposed by the author \cite{M79,Mbk2,book93}
(see also \cite{RPIothers}) following an idea of R.Feynman
\cite{Feynman48}. A Lindblad-type master equation for the density
matrix of an open measured system can be derived from this approach
\cite{master-eq}.  Analogous equations follow from concrete models of
continuous quantum measurements
\cite{CaldeiraLegg83,Milburn88,KMNamiot}.

To describe monitoring a quantum observable in the framework of the
RPI approach, one has to define {\em quantum corridors}, corresponding
to different readouts of the measurement. In the preceding works the
quantum corridors were used which corresponded to the assumption that
monitoring is performed with the ideal resolution of time. In the
present paper we shall consider a more general definition of quantum
corridors including the effect of a {\em finite resolution of time}.
The evolution of a measured system is presented in the resulting
theory by an influence functional, but it cannot be described by a
differential equation (for example a master equation).

Finally the results of the phenomenological consideration will be
compared with conclusions based on a model. For this goal a
modification of the model \cite{KMNamiot} will be presented which
allows one to take into account a finite resolution of time. The
description of the measured system following from the modified model
will be shown to agree with the conclusions of the RPI approach.

\section{Quantum corridors}

A measured system is considered in the RPI theory of continuous
measurements as an open system. The back influence of a measuring
device (environment) onto the measured system is taken into account by
restricting the Feynman path integral presenting the propagator. The
restriction is determined by the information about the measured system
supplied by the measurement. Let us outline this approach (see
\cite{book93} for details).

The evolution of a {\em closed} quantum system during a time interval
$T$ is described by the evolution operator $U_T$.
A matrix element of this operator between the states with
definite positions (in the configuration space) is called the
{\em propagator} and may be expressed in the form of the Feynman path
integral (the variables $q$ and $p$ may be multidimensional)
\be
U_T(q'',q')=\la q''|U_T|q'\ra=\int_{q'}^{q''} \dd [p,q]\,
\exp\left[
\frac{i}{\hbar}\int_0^T (p\dot q - H(p,q,t))
\right].
\label{Feyn}\ee
If the system undergoes a continuous (prolonged in time) measurement
and therefore is considered as being {\em open}, its evolution may be
described (in the RPI approach) by the set of {\em partial evolution
operators} $U_T^{\alpha}$ depending on outputs (readouts) $\alpha$
of the measurement:
\be
|\psi_T^{\alpha}\rangle = U_T^{\alpha} |\psi_0\rangle, \quad
\rho_T^{\alpha} = U_T^{\alpha} \rho_0 \left(
U_T^{\alpha}\right)^{\dagger}.
\label{evolut}\ee
The {\em partial propagators} are expressed by restricted path
integrals. This means that the path integral for
$U_T^{\alpha} $ must be of the form (\ref{Feyn}) but with the
integration restricted according to the information given by the
measurement readout $\alpha$. The information given by $\alpha$ may be
presented by a weight functional $w_\alpha[p,q]$ (positive, with
values between 0 and 1) so that the partial propagator has to be
written as a weighted path integral
\be
U_T^{\alpha}(q'',q') =
\la q''|U_T^{\alpha}|q'\ra=
\int_{q'}^{q''} \dd [p,q]\,w_\alpha[p,q]\,
\exp\left[
\frac{i}{\hbar}\int_0^T (p\dot q - H(p,q,t))
\right].
\label{RPIweight}\ee
The probability density for $\alpha$ to arise as a measurement readout
is given by the trace of the density matrix $\rho_T^{\alpha}$ so that
the probability for  $\alpha$ to belong to some set $\cal A$ of
readouts is
\be
{\rm {Prob}}(\alpha\in{\cal A})=
\int_{{\cal A}} \dd \alpha\, {\rm Tr}\, \rho_T^{\alpha}
\label{prob}\ee
with an appropriate measure $\dd \alpha$ on the set of readouts.

The preceding consideration concerns the situation when the
measurement readout $\alpha$ is known (a {\em selective} description
of the measurement). If the readout is unknown (a {\em non-selective}
description), the evolution of the measured system may be presented by
the complete density matrix
\be
\rho_T = \int \dd \alpha\,  \rho_T^{\alpha}
=\int \dd \alpha\, U_T^{\alpha} \rho_0 \left( U_T^{\alpha}\right)^{\dagger}.
\label{dens-matr}\ee
The generalized unitarity condition
\be
\int \dd \alpha\,  \left( U_T^{\alpha}\right)^{\dagger}\, U_T^{\alpha} =
\bf 1
\label{gen-unitar}\ee
provides conservation of probabilities.

In the special case, when {\em monitoring an observable}
$A=A(p,q,t)$ is considered as a continuous measurement,
the measurement readout is given by the curve
$$
[a] = \{a(t)|0\le t \le T\}
$$
characterizing values of this observable in different time moments.
If  the square average deflection
is taken as a measure of the deviation of the observable
$A(t)=A(p(t),q(t),t)$ from the readout $a(t) $, then the weight
functional describing the measurement may be
taken\footnote{The choice of the weight functional depends on
the class of measurements under consideration.} in the
Gaussian form:
\be
w_{[a]}[p,q]
= \exp\left[-\kappa \int_0^T (A(t) - a(t))^2\,\dd t
  \right].
\label{wa}\ee
The constant $\kappa$ characterizes the resolution of the
measurement and may be expressed in terms of the
``measurement error" $\Delta a_T$ achieved during
the period $T$ of the measurement, $\kappa=1/T\Delta a_T^2$.
The error $\Delta a_T$ decreases with the duration $T$ of the
measurement increased, $\Delta a_T\sim1/\sqrt{T}$.

The resulting path integral
\be
U_T^{[a]}(q'',q')=\int_{q'}^{q''} \dd [p,q]
\exp\left\{ \frac{i}{\hbar} \int_0^T \big(p\dot q
- H\big)\dd t  - \kappa \int_0^T \big(A(t)-a(t)\big)^2 \dd t
\right\}
\label{Ua}\ee
has the form of a conventional (non-restricted) Feynman path
integral (\ref{Feyn}) but with the non-Hermitian
{\em effective Hamiltonian}
\be
H_{[a]}(p,q,t) = H(p,q,t) - i\kappa\hbar \,\big( A(p,q,t) - a(t)
\big)^2
\label{effectHam}\ee
instead of the original Hamiltonian $H$. Therefore, the partial
propagator \eq{Ua} satisfies a Schr\"odinger equation with the
effective Hamiltonian.

This allows one to describe a continuous measurement (monitoring)
without calculating a restricted path integral.
Instead, one may solve the Schr\"odinger equation (with the effective
Hamiltonian) for a wave function of the system:
\be
\frac{\partial}{\partial t} |\psi_t\rangle
  = -\frac{i}{\hbar} H_{[a]}
  = \left(-\frac{i}{\hbar} H
  -\kappa \,\Big( A - a(t)\Big) ^2\right)\, |\psi_t\rangle.
\label{eq-gen}\ee
If the initial wave function $\psi_0$ corresponds to
the initial state of the measured system, then the
solution $\psi_T$ in the final time moment presents the state of
the system after the measurement, under the condition that the
measurement readout is $[a]$.

The wave function $\psi_T$ obtained in this way has a non-unit norm.
If the initial wave function is normalized, then the norm of the final
wave function, according to \Eq{prob}, determines the probability
density of the measurement output: $P[a]=||\psi_T||^2$. Solving the
Schr\"odinger equation for the same initial condition but for
different readouts $[a]$, one has a probability distribution over all
possible scenarios of the measurement with the corresponding final
states of the measured system.

The {\em non-selective description} of the measurement (if the
readout is unknown) is given by the
density matrix $\rho_t$ defined by (\ref{dens-matr}) and satisfying
\cite{master-eq} the equation
\begin{equation}
\dot\rho =-\frac{i}{\hbar}[H,\rho]-\frac{\kappa}{2}[A,[A,\rho]].
\label{non-select}\end{equation}
The influence of the measuring device (measuring medium) on the
measured system may be described by an {\em influence functional}
$W[p,q|p',q']$ (see \cite{Feyn}) in the sense that the
`superpropagator' describing the evolution of the density matrix
\be
\rho_T(q,q')
= \int \dd q_0 \dd q'_0 \; U(q,q'|q_0,q'_0)\rho_0(q_0,q'_0)
\label{superprop}\ee
may be presented in the form of a double path integral:
\ba
\lefteqn{U(q,q'|q_0,q'_0)
=\int_{q}^{q_0} \dd [p,q]\,
\int_{q'}^{q'_0} \dd [p',q']\, W[p,q|p',q']}\nonumber\\
&\times&\exp\left[
\frac{i}{\hbar}\int_0^T [(p\dot q - H(p,q,t))
-(p'\dot q' - H(p',q',t))]
\right]
\label{InflFunct}\ea
In the present case the influence functional may easily be
derived from \Eq{dens-matr} and has the form
\be
W[p,q|p',q'] = \int \dd [a]\,  w_{[a]}[p,q]\, w_{[a]}[p',q'].
\label{W-gen}\ee
With the weight functional \eq{wa} this gives, up to an
inessential number factor,
\be
W[p,q|p',q']=\exp\left[
   -\frac{\kappa}{2}\int_0^T \dd t\,(A(p,q,t) - A(p',q',t))^2
   \right].
\label{W}\ee

\section{Quantum corridors for a finite time resolution}

It has been assumed in the preceding arguments that the time is
measured precisely. This means that the number $a(t)$ is an estimate,
supplied by the measurement, of a value $A(t)$ of the observable
$A$ at the precisely known instant $t$. This assumption is not always
realistic. A real device gives an estimate of the observable $A$ over
some finite time interval of the duration, say, $\tau$. We shall refer
to this situation as ``measuring time with the resolution $\tau$''.

To be more concrete, let $a(t)$ be an estimate, due to the
measurement, of the entity
\be
\A(t) = \la A \ra_t = \int \dd t'\, \Pi_t(t')\,A(t')
\label{av-A}\ee
defined by an appropriate `form-factor' $\Pi_t$ (depending on the type
of the measuring device).
Then, instead of \Eq{wa}, the weight functional for restricting the
path integral should be defined as follows:
\be
\w_{[a]}[p,q] = \exp\left[-\kappa \int_0^T
( a(t) - \A(t) )^2\,\dd t \right].
\label{wa-av}\ee
We arrive therefore, instead of \Eq{Ua}, to the following expression
for the partial propagator:
\be
U_T^{[a]}(q'',q')=\int_{q'}^{q''} \dd [p,q]
\exp\left\{ \frac{i}{\hbar} \int_0^T \big(p\dot q
- H\big)\dd t  - \kappa \int_0^T \big(\A(t)-a(t)\big)^2 \dd t
\right\}.
\label{Ua-av}\ee
It differs radically in that $\A(t)$ depends on $A(t')$ for different
time moments $t'$. Therefore, the description of the evolution is {\em
`not local in time'} and cannot be reduced to an effective Hamiltonian
in analogy with \Eqs{effectHam}{eq-gen}.

The partial propagators \eq{Ua-av} describe an evolution of the
measured system selectively, i.e. with the measurement output $[a]$
taken into account. A non-selective description is given by the
general formula \eq{dens-matr} resulting in the present case in an
influence functional of the form
\ba
\W[p,q|p',q'] &=& \int \dd [a]\,  \w_{[a]}[p,q]\, \w_{[a]}[p',q']\nonumber\\
&=&\exp\left[
   -\frac{\kappa}{2}\int_0^T \dd t\,(\A(p,q,t) - \A(p',q',t))^2
   \right].
\label{W-av}\ea
\begin{remark}
The measure $\dd [a]$ of integration over measurement readouts has to
be chosen in such a way that the generalized unitarity \Eq{gen-unitar}
be valid. \Eq{W-av} is valid with the conventional functional measure
$\dd [a]\sim\prod_t \dd a(t)$ which provides the generalized unitarity
either for a linear measured system or for a system with a not too large
nonlinearity and a measurement with a not too high resolution. I a 
general case a weight has to be included in the measure $\dd [a]$. 
\Eq{W-av} should be modified in this case.  \end{remark}

The influence functional \eq{W-av} enables one to describe the
evolution of the measured system by \Eqs{superprop}{InflFunct}
(with $\W$ instead of $W$), but no differential equation in time
(analogous to \Eq{non-select}) exists for the resulting density
matrix. This is a consequence of non-locality in time.

In the rest of the paper we shall consider a concrete model of the
monitoring the position of a particle to verify that it actually leads
to the influence functional of the form \Eq{W-av}.

\section{Finite resolution in the model of a continuous measurement}
\label{model}

The well-known model of a quantum diffusion proposed by Caldeira and
Leggett \cite{CaldeiraLegg83} may be considered as a model for a
continuous measurement, namely, for monitoring the position of a
particle by a measuring medium. In this model the decoherence
(measurement) is caused by the interaction of the particle with modes
of the crystal.

One more model of this type has been proposed in \cite{KMNamiot}. This
model also consists of a particle in some medium. However the ``atoms"
of the medium are modelled as oscillators not interacting with each
other. Decoherence is caused in this case by interaction of the
particle with internal degrees of freedom of the atoms (presented as
degrees of freedom of the oscillators). Let us remark that in most
cases such a model is more realistic than the Caldeira-Leggett model
because the decoherence due to internal structure of atoms is more
efficient (more fast) than the decoherence by modes of a crystal. Now
we shall modify this model to take into account a finite resolution of
time.

The measuring medium in the model \cite{KMNamiot} consists of atoms in
the nods of a cubic lattice. Each atom is presented as an oscillator.
We shall present each atom  as a family of oscillators of different
frequencies. However let us begin by the case of a single oscillator
with the frequency $\om$ as in \cite{KMNamiot}. The Hamiltonian of
the interaction between the oscillator and the particle is taken in
\cite{KMNamiot} as
\be
H_{\rm int}=\sum_k H_{k}
= \sum_k \gamma_\om q_k \exp\left[
-\frac{(\r-\cc_k)^2}{l^2}
\right].
\label{interact}\ee
Here ${\r}$ is a position of the measured particle while $q_k$ is a
canonical coordinate of the $k$th oscillator, ${\cc}_k$ its location,
$\gamma_\om$ the interaction constant, and the length $l$
characterizes the range of the interaction. The interaction of the
particle with each of the atoms (oscillators) may be considered as a
measurement of its location with the precision $l$.

To solve this model, the discrete lattice of atoms was replaced in
\cite{KMNamiot} by a continuous distribution of them with the constant
density $n$. This enables one to calculate the influence functional
describing the influence of the medium on the particle:
\ba
W[\r_1,\r_2]
&=& \exp\left\{
-\int_0^T\dd t \int_0^T\dd t'\; \nu(\om)\cos\omega(t-t')
\left[\exp\left(
-\frac{(\r_1(t)-\r_1(t'))^2}{2l^2}\right)\right.\right.\nonumber\\
&+& \left.\left.\exp\left( -\frac{(\r_2(t)-\r_2(t'))^2}{2l^2}\right)
-2 \exp\left( -\frac{(\r_2(t)-\r_1(t'))^2}{2l^2}\right) \right]
\right\} \label{influence-om}\ea
where
\be
\nu(\om)={n}\left( \frac{\pi l^2}{2} \right)^{3/2}
\frac{\gamma_\om^2}{4\hbar m\omega}.
\label{nu}\ee

This simple model is not enough realistic because it does not describe
measuring of time: the state of an oscillator after the interaction
with the particle contains no information about the time of the
interaction. To describe the measurement of time, it was assumed in
\cite{KMNamiot} that the interaction of each oscillator is turned on
for a short time. Let us consider now a more realistic model for the
measurement of time.

For this goal, assume that each atom has a more complicated internal
structure so that the information about the time of the interaction is
recorded in its state. To give a model of the internal structure of
the atom, let us present it as a family of oscillators with different
frequencies. Now the information about the time of interaction with
the particle is recorded in phase relations between different
oscillators of the same `atom'.

Going over to the calculation, we have now integrate the exponent in
\Eq{influence-om} over the frequencies $\om$ of the oscillators
forming the model of an atom. The influence functional takes the form
\ba
\lefteqn{W[\r_1,\r_2]
= \exp\left\{
-\frac{\k l^2}{2}\int_0^T\dd t \int_0^T\dd t'\; \Pi(t-t')
\left[\exp\left(
-\frac{(\r_1(t)-\r_1(t'))^2}{2l^2}\right)\right.\right.}\nonumber\\
&&+ \left.\left.\exp\left( -\frac{(\r_2(t)-\r_2(t'))^2}{2l^2}\right)
-2 \exp\left( -\frac{(\r_2(t)-\r_1(t'))^2}{2l^2}\right) \right]
\phantom{\int_0^T}\hspace{-2ex}
\right\}
\phantom{xxx}
\label{influence-time}\ea
with
\be
\Pi(t)=\frac{2}{\k l^2}\int  \nu(\om)\cos\omega t \dd\om,
\quad
\int \Pi(t)\,\dd t = 1.
\label{form-factor}
\ee

Let the width of the `form-factor' $\Pi(t)$ be of the order of
$\tau$. Then for not too large initial energy of the particle,
$E_0\ll {Ml^2}/{2\tau^2}$, the exponentials in the square brackets may
be evaluated up to the first order to give
\ba
\lefteqn{W[\r_1,\r_2]
= \exp\left\{
-\frac{\k}{4}\int_0^T\dd t' \int_0^T\dd t''\; \Pi(t'-t'')\,
\right.}\nonumber\\
&&\left.
[2 (\r_2(t')-\r_1(t''))^2
-(\r_1(t')-\r_1(t''))^2
- (\r_2(t')-\r_2(t''))^2]\phantom{\int_0^T}\hspace{-2ex}
\right\} \phantom{xxx}
\label{influence-approx}\ea

Let the function $\Pi(t)$ be symmetrical. Then it may be expressed
through the function of two arguments $\Pi_t(t')$ (depending only
on the difference $|t-t'|$):
\be
\Pi(t'-t'')=\int \dd t\, \Pi_{t}(t') \, \Pi_{t}(t''), \quad
\int \Pi_{t}(t')\dd t'=1.
\label{Pi-Pi0}\ee
Let us substitute this expression in
\Eq{influence-approx} and make use of the relation
\ba
\lefteqn{\int \dd t'\int \dd t''\, \Pi_{t}(t') \, \Pi_{t}(t'') }\nonumber\\
&\times&[2(\r_2(t')-\r_1(t''))^2
     -(\r_1(t')-\r_1(t''))^2-(\r_2(t')-\r_2(t''))^2]\nonumber\\
&=&2\int \dd t[\la\r_2\ra_{t} - \la\r_1\ra_{t}]^2
\label{r2}\ea
where the notation \eq{av-A} is exploited. Then we immediately see
that the influence functional has the form
\be
W[\r,\r']=\exp\left[
   -\frac{\kappa}{2}\int_0^T \dd t\,(\la \r\ra_{t} - \la \r'\ra_{t})^2
   \right]
\label{Wr-av}\ee
in accord with \Eq{W-av}. Thus, the prediction of the phenomenological
RPI approach coincides with what follows from the concrete model of
the measurement even in the case when a finite resolution of time is
taken into account.

\section{Conclusion}

We have shown, both in the RPI phenomenological approach and in the
framework of the concrete model of a continuous measurement, that the
finiteness of the resolution in measuring time moments leads to the
replacement of the value of an observable $A(t)$ by its
`time-coarse-graining' $\A(t)=\la A\ra_t$ in the expression for the
influence functional. In a special case when the position $\r(t)$ of a
particle is continuously measured, $\r(t)$ has to be replaced by $\la
\r\ra_t$. As a result, the time evolution of the measured system
cannot be presented by a differential equation in time. Neither the
master equation \Eq{non-select} nor the Schr\"odinger equation with
a complex Hamiltonian \Eq{eq-gen} is correct in this case.

Time coarse-graining is a characteristic of the `measuring device' or
`measuring medium'. It characterizes inertial properties of the
measuring setup. It is evident that the effect of time coarse-graining
is negligible if all physically relevant frequencies are lower than
the inverse period of coarse-graining: $\Omega\ll\tau^{-1}$. In this
case the master equation and the Schr\"odinger equation with a complex
Hamiltonian are correct. Physically this means that inertial
properties of the measuring medium are negligible.

\vspace{0.5cm}
\centerline{ACKNOWLEDGEMENT}

The author is indebted to V.Namiot for stimulating
discussions. This work was supported in part by the Deutsche
Forschungsgemeinschaft.


\newpage
\begin{thebibliography}{10}

\bibitem{cont-meas-model} H.D.Zeh, Found. Phys. {\bf 1}, 69 (1970);
V.B.Braginsky and Yu.I.Vorotsov,
Usp. Fiz. Nauk {\bf 114}, 41 (1974) [Sov. Phys. Usp, {\bf 17}, 644
(1975)];
M.D.Srinivas, {\em J. Math. Phys.} {\bf 18}, 2138 (1977);
D.F.Walls and G.J.Milburn, Phys. Rev. {\bf A 31}, 2403 (1985);
E.Joos, and H.D.Zeh, Z.Phys. {\bf B 59}, 223 (1985);
A.Peres, {\em Quantum Theory: Concepts and Methods},
Kluwer Academic Publishers, Dordrecht, Boston \& London, 1993;
H.Carmichael, {\em An Open Systems Approach to Quantum Optics},
Springer Verlag, Berlin, Heidelberg, 1993;
D.Giulini, E.Joos, C.Kiefer, J.Kupsch, I.-O.Stamatescu and H.D.Zeh,
{\em Decoherence and the appearance of a classical world in quantum
theory}, Springer, Berlin etc., 1996.

\bibitem{CaldeiraLegg83} A.O.Caldeira and A.J.Leggett,
Path integral approach to quantum Brownian motion,
Physica {\bf A~121}, 587-616 (1983).

\bibitem{Milburn88} G.J.Milburn, J. Opt. Soc. Am. {\bf B 5}, 1317
(1988).

\bibitem{KMNamiot} A.Konetchnyi, M.B.Mensky and V.Namiot, Phys.
Lett. {\bf A 177}, 283 (1993).

%%% Contin. Meas. phenomenology
%%% {cont-meas-phenomen,M79,Mbk2,book93}

\bibitem{cont-meas-phenomen}
G.Lindblad,
Commun. Math. Phys. {\bf 48}, 119 (1976);
E.B.Davies,
{\em Quantum Theory of Open Systems},
Academic Press, London, New York, San Francisco, 1976;
L.Diosi, Phys. Lett. {\bf A 129}, 419 (1988);
N.Gisin,
Helv. Phys. Acta {\bf 62}, 363-371 (1989);
C.Presilla, R.Onofrio and U.Tambini,
Ann. Phys. (USA) {\bf 248}, 95-121 (1996).
%%% RPI

\bibitem{M79} M.B.Mensky, Phys. Rev. {\bf D 20}, 384 (1979);
Sov. Phys.-JETP {\bf 50}, 667 (1979).

\bibitem{Mbk2} M.B.Mensky, {\em The Path Group: Measurements, Fields,
Particles} (Nauka, Moscow, 1983, in Russian; Japan translation:
Yosioka, Kyoto, 1988).

\bibitem{book93} M.B.Mensky, {\em Continuous Quantum
Measurements and Path Integrals}, IOP Publishing, Bristol and
Philadelphia, 1993.

%%% Zeno effect: {Zeno,Itano}

\bibitem{Zeno} B.Misra and E.C.G.Sudarshan, J. Math. Phys. {\bf
18}, 756 (1977); C.B.Chiu, E.C.G.Sudarshan and B.Misra,
Phys. Rev. {\bf D 16}, 520 (1977); A.Peres, Amer. J. Phys. {\bf
48}, 931 (1980).

\bibitem{Itano} W.M.Itano, D.J.Heinzen, J.J.Bollinger and
D.J.Wineland, Phys. Rev. {\bf A 41}, 2295 (1990).

%% \bibitem{Khal81} \bibitem{Barch82} \bibitem{Caves86/87}

\bibitem{RPIothers}
F.Ya.Khalili, Vestnik Mosk. Universiteta, Ser.~3,
Phys., Astr., {\bf 22}, 37 (1981);
A.Barchielli, L.Lanz and G.M.Prosperi, Nuovo
Cimento, {\bf B 72}, 79 (1982);
C.M.Caves, Phys. Rev. {\bf D33}, 1643 (1986);
Phys. Rev. {\bf D35}, 1815 (1987).

%%% Feynman

\bibitem{Feynman48}
R.P.Feynman, Rev. Mod. Phys. {\bf 20}, 367 (1948).

%%% MBM other

\bibitem{master-eq} M.B.Mensky, Phys. Lett. {\bf
A 196}, 159 (1994).

\bibitem{Feyn} R.P.Feynman and F.L.Vernon, Ann. Phys. {\bf 24},
118 (1963);
R.Feynman, A.Hibbs,
{\em Quantum mechanics and path integrals},
McGraw-Hill Book Co., New York, 1965.

\end{thebibliography}
\end{document}